# Security and Privacy Issues of Big Data

**José Moura[1,2], Carlos Serrão[1]**

[1] ISCTE-IUL, Instituto Universitário de Lisboa, Portugal

[2] IT, Instituto de Telecomunicações, Lisboa, Portugal

{jose.moura, carlos.serrao}@iscte.pt

## ABSTRACT

*This chapter revises the most important aspects in how computing infrastructures should be configured and intelligently managed to fulfill the most notably security aspects required by Big Data applications. One of them is privacy. It is a pertinent aspect to be addressed because users share more and more personal data and content through their devices and computers to social networks and public clouds. So, a secure framework to social networks is a very hot topic research. This last topic is addressed in one of the two sections of the current chapter with case studies. In addition, the traditional mechanisms to support security such as firewalls and demilitarized zones are not suitable to be applied in computing systems to support Big Data. SDN is an emergent management solution that could become a convenient mechanism to implement security in Big Data systems, as we show through a second case study at the end of the chapter. This also discusses current relevant work and identifies open issues.*

## INTRODUCTION

The Big Data is an emerging area applied to manage datasets whose size is beyond the ability of commonly used software tools to capture, manage, and timely analyze that amount of data. The quantity of data to be analyzed is expected to double every two years (IDC, 2012). All these data are very often unstructured and from various sources such as social media, sensors, scientific applications, surveillance, video and image archives, Internet search indexing, medical records, business transactions and system logs. Big data is gaining more and more attention since the number of devices connected to the so-called "Internet of Things" (IoT) is still increasing to unforeseen levels, producing large amounts of data which needs to be transformed into valuable information. Additionally, it is very popular to buy on-demand additional computing power and storage from public cloud providers to perform intensive data-parallel processing. In this way, security and privacy issues can be potentially boosted by the volume, variety, and wide area deployment of the system infrastructure to support Big Data applications.

As Big Data expands with the help of public clouds, traditional security solutions tailored to private computing infrastructures, confined to a well-defined security perimeter, such as firewalls and demilitarized zones (DMZs) are no more effective. Using Big Data, security functions are required to work over the heterogeneous composition of diverse hardware, operating systems, and network domains. In this puzzle-type computing environment, the abstraction capability of Software-Defined Networking (SDN) seems a very important characteristic that can enable the efficient deployment of Big Data secure services on-top of the heterogeneous infrastructure. SDN introduces abstraction because it separates the control (higher) plane from the underlying system infrastructure being supervised and controlled. Separating a network's control logic from the underlying physical routers and switches that forward traffic allows system administrators to write high-level control programs that specify the behavior of an entire network, in contrast to conventional networks, whereby administrators (if allowed to do it by the device manufacturers) must codify functionality in terms of low-level device configuration. Using SDN, the intelligent management of secure functions can be implemented in a logically centralized controller, simplifying the following aspects: implementation of security rules; system (re)configuration; and system evolution. The robustness drawback of a centralized SDN solution can be mitigated using a hierarchy of controllers and/or through the usage of redundant controllers at least for the most important system functions to be controlled.

The National Institute of Standards and Technology (NIST) launched very recently a framework with a set of voluntary guidelines to help organizations make their communications and computing operations safer (NIST, 2014). This could be achieved through a systematic verification of the system infrastructure in terms of risk assessment, protection against threats, and capabilities to respond and recover from attacks. Following the last verification principles, Defense Advanced Research Projects Agency (DARPA) is creating a program called Mining and Understanding Software Enclaves (MUSE) to enhance the quality of the US military's software. This program is designed to produce more robust software that can work with big datasets without causing errors or crashing under the sheer volume of information (DARPA, 2014). In addition, security and privacy are becoming very urgent Big Data aspects that need to be tackled (Agrawal, Das, & El Abbadi, 2011). To illustrate this, the social networks have enabled people to share and distribute valuable copyrighted digital contents in a very easy way. Consequently, the copyright infringement behaviors, such as illicit copying, malicious distribution, unauthorized access and usage, and free sharing of copyright-protected digital contents, will become a much more common phenomenon. To mitigate these problems, Big Data should have solid solutions to support author's privacy and author's copyrights (Marques & Serrão, 2013a). Also, users share more and more personal data and user generated content through their mobile devices and computers to social networks and cloud services, loosing data and content control with a serious impact on their own privacy. Finally, one potentially promising approach is to create additional uncertainty for attackers by dynamically changing system properties in what is called a cyber moving target (MT) (Okhravi, Hobson, Bigelow, & Streilein, 2014). They present a summary of several types of MT techniques, consider the advantages and weaknesses of each, and make recommendations for future research in this area.

The current chapter endorses the most important aspects of Big Data security and privacy and is structured as follows. The first section discusses the most important challenges to the aspects of information security and privacy imposed by the novel requirements of Big Data applications. The second section presents and explains some interesting solutions to the problems found in the previous section. The third and fourth sections are related with two case studies in this exciting emergent area.

## BIG DATA CHALLENGES TO INFORMATION SECURITY AND PRIVACY

With the proliferation of devices connected to the Internet and connected to each other, the volume of data collected, stored, and processed is increasing everyday, which also brings new challenges in terms of the information security. In fact, the currently used security mechanisms such as firewalls and DMZs cannot be used in the Big Data infrastructure because the security mechanisms should be stretched out of the perimeter of the organization's network to fulfill the user/data mobility requirements and the policies of BYOD (Bring Your Own Device). Considering these new scenarios, the pertinent question is what security and privacy policies and technologies are more adequate to fulfill the current top Big Data privacy and security demands (Cloud Security Alliance, 2013). These challenges may be organized into four Big Data aspects such as infrastructure security (e.g. secure distributed computations using MapReduce), data privacy (e.g. data mining that preserves privacy/granular access), data management (e.g. secure data provenance and storage) and, integrity and reactive security (e.g. real time monitoring of anomalies and attacks).

Considering Big Data there is a set of risk areas that need to be considered. These include the information lifecycle (provenance, ownership and classification of data), the data creation and collection process, and the lack of security procedures. Ultimately, the Big Data security objectives are no different from any other data types – to preserve its confidentiality, integrity and availability.

Being Big Data such an important and complex topic, it is almost natural that immense security and privacy challenges will arise (Michael & Miller, 2013; Tankard, 2012). Big Data has specific characteristics that affect information security: variety, volume, velocity, value, variability, and veracity (Figure 1). These challenges have a direct impact on the design of security solutions that are required to tackle all these characteristics and requirements (Demchenko, Ngo, Laat, Membrey, & Gordijenko, 2014). Currently, such out of the box security solution does not exist.

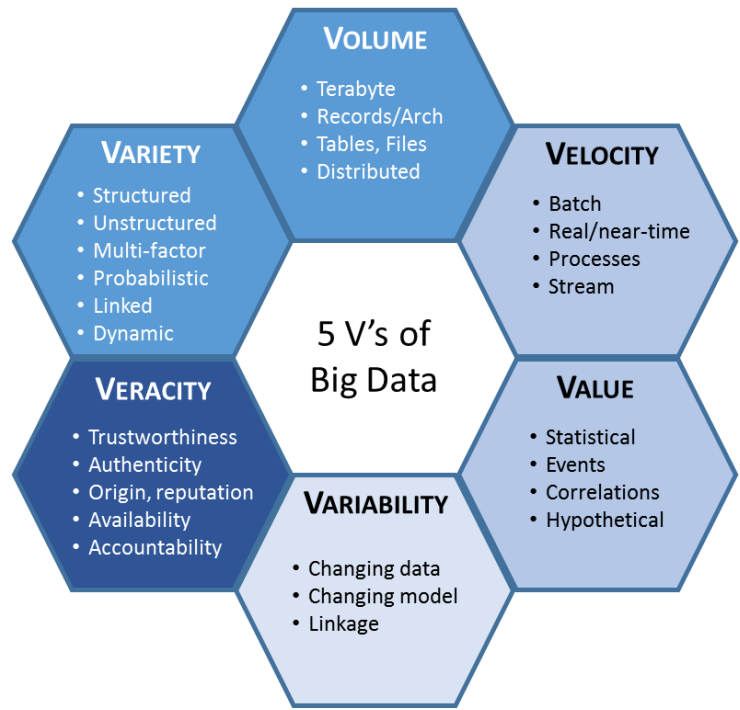

Figure 1. The five V's of Big Data (adapted from ("IBM big data platform - Bringing big data to the Enterprise," 2014))

Cloud Secure Alliance (CSA), a non-profit organization with a mission to promote the use of best practices for providing security assurance within Cloud Computing, has created a Big Data Working Group that has focused on the major challenges to implement secure Big Data services (Cloud Security Alliance, 2013). CSA has categorized the different security and privacy challenges into four different aspects of the Big Data ecosystem. These aspects are Infrastructure Security, Data Privacy, Data Management and, Integrity and Reactive Security. Each of these aspects faces the following security challenges, according to CSA:

- Infrastructure Security
    1. Secure Distributed Processing of Data
    2. Security Best Actions for Non-Relational Data-Bases
- Data Privacy
    3. Data Analysis through Data Mining Preserving Data Privacy
    4. Cryptographic Solutions for Data Security
    5. Granular Access Control
- Data Management and Integrity
    6. Secure Data Storage and Transaction Logs
    7. Granular Audits
    8. Data Provenance
- Reactive Security

9. End-to-End Filtering & Validation
10. Supervising the Security Level in Real-Time

These security and privacy challenges cover the entire spectrum of the Big Data lifecycle (Figure 2): sources of data production (devices), the data itself, data processing, data storage, data transport and data usage on different devices.

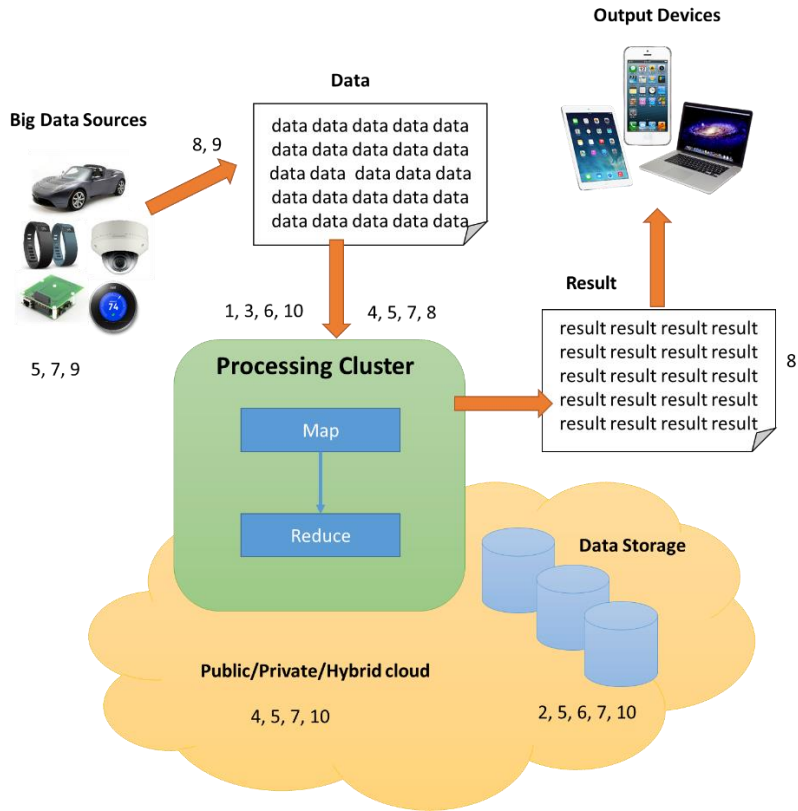

Figure 2. Security and Privacy challenges in Big Data ecosystem (adapted from (Cloud Security Alliance, 2013))

A particular aspect of Big Data security and privacy has to be related with the rise of the Internet of Things (IoT). IoT, defined by Oxford[1] as "a proposed development of the Internet in which everyday objects have network connectivity, allowing them to send and receive data", is already a reality – Gartner estimates that 26 billion of IoT devices will be installed by 2020, generating an incremental revenue of $300 billion (Rivera & van der Meulen, 2014). The immense increase in the number of connected devices (cars, lighting systems, refrigerators, telephones, glasses, traffic control systems, health monitoring devices, SCADA systems, TVs, home security systems, home automation systems, and many more) has led to manufacturers to push to the market, in a short period of time, a large set of devices, cloud systems and mobile applications to exploit this opportunity. While it presents tremendous benefits and opportunities for end-users it also is responsible for security challenges.

HP recently conducted a study on market-available IoT solutions and concluded that 70% of those contain security problems. These security problems were related with privacy issues, insufficient authorization, lack of transport encryption, insecure web interface and inadequate software protection (HP, 2014). Based on some of these findings, HP has started a project at OWASP (Open Web Application Security Project) that is entitled "OWASP Internet of Things Top Ten" (OWASP, 2014) whose objective is to help IoT suppliers to identify the top ten security IoT device problems and how to avoid them. This project, similar to the OWASP Top 10, identified the following security problems:

- Insecure Web Interface: which can allow an attacker to exploit an administration web interface (through cross-site scripting, cross-site request forgery and SQL injection) and obtain unauthorized access to control the IoT device.
- Insufficient Authentication/Authorization: can allow an attacker to exploit a bad password policy, break weak passwords and access to privileged modes on the IoT device.
- Insecure Network Services: which can lead to an attacker exploiting unnecessary or weak services running on the device, or use those services as a jumping point to attack other devices on the IoT network.
- Lack of Transport Encryption: allowing an attacker to eavesdrop data in transit between IoT devices and support systems.
- Privacy Concerns: raised from the fact the most IoT devices and support systems collect personal data from users and fail to protect that data.
- Insecure Cloud Interface: without proper security controls an attacker can use multiple attack vectors (insufficient authentication, lack of transport encryption, account enumeration) to access data or controls via the cloud website.
- Insecure Mobile Interface: without proper security controls an attacker can use multiple attack vectors (insufficient authentication, lack of transport encryption, account enumeration) to access data or controls via the mobile interface.
- Insufficient Security Configurability: due to the lack or poor configuration mechanisms an attacker can access data or controls on the device.
- Insecure Software/Firmware: attackers can take advantage of unencrypted and unauthenticated connections to hijack IoT devices updates, and perform malicious update that can compromise the device, a network of devices and the data they hold.
- Poor Physical Security: if the IoT device is physically accessible than an attacker can use USB ports, SD cards or other storage means to access the device OS and potentially any data stored on the device.

---
[1] http://www.oxforddictionaries.com/definition/english/Internet-of-things

It is clear that Big Data present interesting opportunities for users and businesses, however these opportunities are countered by enormous challenges in terms of privacy and security (Cloud Security Alliance, 2013). Traditional security mechanisms are insufficient to provide a capable answer to those challenges. In the next section, some of these solutions/proposals are going to be addressed.

## SOLUTIONS/PROPOSALS TO ADDRESS BIG DATA SECURITY AND PRIVACY CHALLENGES

There is no single magical solution to solve the identified Big Data security and privacy challenges and traditional security solutions, which are mainly dedicated to protect small amounts of static data, are not adequated to the novel requisites imposed by Big Data services (Cloud Security Alliance, 2013). There is the need to understand how the collection of large amounts of complex structured and unstructured data can be protected. Non-authorized access to that data to create new relations, combine different data sources and make it available to malicious users is a serious risk for Big Data. The basic and more common solution for this includes encrypting everything to make data secure regardless where the data resides (data center, computer, mobile device, or any other). As Big Data grows and its processing gets faster, then encryption, masking and tokenization are critical elements for protecting sensitive data.

Due to its characteristics, Big Data projects need to take an holistic vision at security (Tankard, 2012). Big Data projects need to take into consideration the identification of the different data sources, the origin and creators of data, as well as who is allowed to access the data. It is also necessary to conduct a correct classification to identify critical data, and align with the organization information security policy in terms of enforcing access control and data handling policies. As a recommendation, different security mechanisms should be closer to the data sources and data itself, in order to provide security right at the origin of data, and mechanisms of control and prevention on archiving, data leakage prevention and access control should work together (Kindervag, Balaouras, Hill, & Mak, 2012).

The new Big Data security solutions should extend the secure perimeter from the enterprise to the public cloud (Juels & Oprea, 2013). In this way, a trustful data provenance mechanism should be also created across domains. In addition, similar mechanisms to the ones used in (Luo, Lin, Zhang, & Zukerman, 2013) can be used to mitigate distributed denial-of-service (DDoS) attacks launched against Big Data infrastructures. Also, a Big Data security and privacy is necessary to ensure data trustworthiness throughout the entire data lifecycle – from data collection to usage.

The personalization feature of some Big Data services and its impact on the user privacy is discussed in (Hasan, Habegger, Brunie, Bennani, & Damiani, 2013). They discuss these issues in the backdrop of EEXCESS, a concrete project aimed to both provide high level recommendations and to respect user privacy. A recent work describes proposed privacy extensions to UML to help software engineers to quickly visualize privacy requirements, and design them into Big Data applications (Jutla, Bodorik, & Ali, 2013).

While trying to take the most of Big Data, in terms of security and privacy, it becomes mandatory that mechanisms that address legal requirements about data handling, need to be met. Secure encryption technology must be employed to protect all the confidential data (Personally Identifiable Information (PII), Protected Health Information (PHI) and Intellectual Property (IP) and careful cryptographic material (keys) access management policies, need to be put in place, to ensure the correct locking and unlocking of data – this is particularly important for data stored. In order to be successful these mechanisms need to be transparent to the end-user and have low impact of the performance and scalability of data (software and hardware-based encryptions mechanisms are to be considered) (Advantech, 2013).

As previously referred, traditional encryption and anonymization of data are not adequate to solve Big Data problems. They are adequate to protect static information, but are not adequate when data computation is involved (MIT, 2014). Therefore, other techniques, allowing specific and targeted data computation while keeping the data secret, need to be used. Secure Function Evaluation (SFE) (Lindell & Pinkas, 2002), Fully Homomorphic Encryption (FHE) (Gentry, 2009) and Functional Encryption (FE) (Goldwasser et al., 2014), and partition of data on non-communicating data centers, can help solving the limitations of traditional security techniques.

Homomorphic encryption is a form of encryption which allows specific types of computations (e.g. RSA public key encryption algorithm) to be carried out on ciphertext and generate an encrypted result which, when decrypted, matches the result of operations performed on the plaintext (Gentry, 2010). Fully homomorphic encryption has numerous applications, as referred in (Van Dijk, Gentry, Halevi, & Vaikuntanathan, 2010). This allows encrypted queries on databases, which keeps secret private user information where that data is normally stored (somewhere in the cloud – in the limit an user can store its data on any untrusted server, but in encrypted form, without being worried with the data secrecy) (Ra Popa & Redfield, 2011). It also enables private queries to a search engine - the user submits an encrypted query and the search engine computes a succinct encrypted answer without ever looking at the query in the clear which could contain private user information such as the number of the national healthcare service. The homomorphic encryption also enables searching on encrypted data - a user stores encrypted files on a remote file server and can later have the server retrieve only files that (when decrypted) satisfy some boolean constraint, even though the server cannot decrypt the files on its own. More broadly, the fully homomorphic encryption improves the efficiency of secure multiparty computation.

An important security and privacy challenge for Big Data is related with the storage and processing of encrypted data. Running queries against an encrypted database is a basic security requirement for secure Big Data however it is a challenging one. This raises questions such as a) is the database encrypted with a single or multiple keys; b) does the database needs to be decrypted prior to running the query; c) do the queries need to be also encrypted; d) who as the permissions to decrypt the database; and many more. Recently a system that was developed at MIT, provides answers to some of these questions. CryptDB allows researchers to run database queries over encrypted data (Ra Popa & Redfield, 2011). Trustworthy applications that intent to query encrypted data will pass those queries to a CryptDB proxy (that sits between the application and the database) that rewrites those queries in a specific way so that they can be run against the encrypted database. The database returns the encrypted results back to the proxy, which holds a master key and will decrypt the results, sending the final answer back to the application. CryptDB supports numerous forms of encryption schemes that allow different types of operations on the data (RA Popa & Redfield, 2012). Based on CryptDB, Google has developed the Encrypted Big Query Client that will allow encrypted big queries against their BigQuery service that enables super, SQL-like queries against append-only tables, using the processing power of Google's infrastructure (Google, 2014).

Apart from more specific security recommendations, it is also important to consider the security of the IT infrastructure itself. One of the common security practices is to place security controls at the edge of the networks however, if an attacker violates this security perimeter it will have access to all the data within it. Therefore, a new approach is necessary to move those security controls near to the data (or add additional ones). Monitoring, analyzing and learning from data usage and access is also an important aspect to continuously improve security of the data holding infrastructure and leverage the already existing security solutions (Kindervag et al., 2012; Kindervag, Wang, Balaouras, & Coit, 2011).

## A CASE STUDY IN A SECURE SOCIAL APPLICATION

Social networks are one of the key-applications for a large number of users. Millions and millions of persons are connected to some kind of social network – e.g. Facebook according to its own accounting has more than 829 million daily active users on average (654 million with mobile access). Social networks are quite attractive to users because they allow communication with new persons and concede users the ability to expose their own network of friends to others, creating new relations and pairings among users and between users and content (McKenzie et al., 2012). Users take advantage of this functionality to share all kinds of digital content within the social network, with other users (either they are their direct contacts or they are in other one's connections). These social network-sharing functionalities are extremely powerful and engaging of further social interaction. However, they are at the same time, the cause of serious privacy and security problems because sharing control is not on the end-user side. This represents a serious threat to the user privacy since content shared in these platforms can easily be exposed to a wider audience in just a few seconds. It is difficult, for an ordinary user, to select specific sharing properties for the content placed in social network and ensure that it stays under its control.

With the emergence of Web 2.0, users have changed from being simply information consumers to become important content producers. User generated content is content that is voluntarily developed by an individual or a consortium, and distributed through an online platform. The volume of user generated content currently produced and made available through several platforms is already immense and continues to grow in size (Kim, Jin, Kim, & Shin, 2012). For instance, Facebook stores, accesses, and analyzes more than 30 Petabytes of user generated data (100 terabytes of data are uploaded daily to Facebook) and YouTube users upload 48 hours of new video every minute of the day (McKenzie et al., 2012).

Currently, social network platforms already present a set of pre-defined but limited content privacy and security sharing controls (X. Chen & Shi, 2009). Major social network platforms offer the possibility for users to share content under specific privacy rules, which are defined by the social platform and not by the end-user. Most of the times, these rules are extremely permissive and differ from platform to platform. Also, on social networks, content is shared in a non-protected manner, making it easier for unauthorized usage and sharing. Users are also bound by subsequent privacy policies changes that threaten more and more the user right to protect its personal information and personal content.

The other problem that is most of the times associated with the security and privacy of content shared on social networks, is related to the security of the social network platform itself (Gross & Acquisti, 2005). The exploitation of the social network infrastructure can lead to security and privacy threats. On the other hand, recently on the media there have been some allegations about the cooperation of some of the most important IT suppliers (including some major social platforms) with governmental agencies to allow the unauthorized access to user's information and content. This latter fact is quite relevant, because, in theory, the social network service supplier has unlimited access to the information and content of all its customers.

This is an increasing serious problem, not only for end-users but also for organizations. More and more, organizations rely on social network services as a mean to disseminate information, create relations with and between employees and customers, knowledge capture and dissemination. The privacy and security challenges presented by these new ways of communication and interaction are very pertinent topics for both end users and organizations.

The continuous growing proliferation of mobile devices (mostly smartphones and tablets, but soon more devices will enter this scenario) with capabilities of producing content (mainly audio recordings, videos and pictures) at the palm of every user's hand, following them everywhere and anytime is also a serious threat to their content privacy and security (De Cristofaro, Soriente, Tsudik, & Williams, 2012). This user generated content creates cultural, symbolic, and affective benefit including personal satisfaction, enhanced skill or reputation, improved functionality for existing games or devices, community building or civic engagement. In more simplistic terms, user generated content creates value, economic or not.

Having all of this into consideration, it seems clear that it is necessary to have a clear separation among the social network platform providers, their social functionalities, and the user generated content that they hold. It is important to create mechanisms that transfer part of the security and sharing control to the end-user side. Having this into consideration, in this section, it is proposed and presented a paradigm shift that implies a change from the current social networks security and privacy scenario based on a social network platform centric, to another paradigm that empowers social networks users' on the control and safeguard of its privacy, passing the user generated content sharing control to the end-user side, using rights management systems (Marques & Serrão, 2013b). Also, the entity that is responsible for the storage and protection of the user generated content is independent of the social network platform itself.

This new approach creates a mechanism that protects the shared user generated content on the social network platform while it provides the content sharing and access control to the end-user.

## Overall system architecture

As referred on the previous section, the novel approach that is followed is based on open rights management systems – in particular, and for this sake, it is based on OpenSDRM (Carlos Serrão, Neves, Trevor Barker, & Massimo Balestri, 2003; Serrão, 2008). OpenSDRM is an open and distributed rights management architecture that allows the implementation of different content business models. Moreover, OpenSDRM was created having into consideration interoperability aspects (Serrão, Rodriguez, & Delgado, 2011) that permit that the different modules that compose the system to be decoupled and re-integrated to allow interoperability (Serrão, Dias, & Kudumakis, 2005; Serrão et al., 2011) with other non-OpenSDRM components, using an open and well-defined API (Figure 3). Additionally there may exist also more than one instance of each of the services on the platform, allowing the scalability and growth of the set of all possible configuration options (Serrão, Dias, & Delgado, 2005).

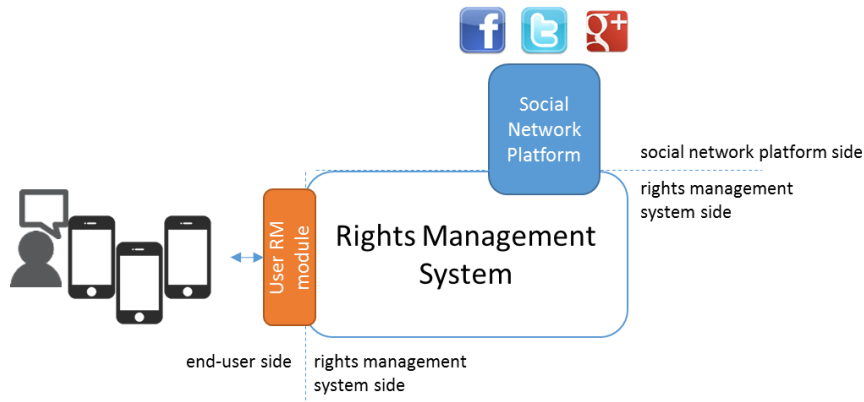

Figure 3 - Overview of the architecture integrated with the rights management system

For the proposed scenario, the social network platform can be integrated with the rights management system, using different methods. If the social network implements a development API or if it is open-source, a much tighter integration scenario can be achieved. If not, it is possible to use other publicly available mechanisms on the platform (or out of the platform) to enable a lesser integrated scenario, but that maintains the privacy and security characteristics sought. Using mechanisms on the platform is the most common scenario and therefore is the approach that will be reflected here.

In this architecture there are some elements that cooperate in order to provide the necessary functionalities to both the end-users and the social network platform, in order to implement the necessary mechanisms to provide security and privacy to user generated content.

OpenSDRM, as an open rights management framework is composed by different services (Figure 4). Some of the services are deployed on the server-side while other are implemented on the user-side. On the user-side, the authorization service handles the requests to render some type of content on the user device, processing the requests and matching them to existing credentials, licenses and permissions to render the content. Also, on the end-user side the content rendering service is responsible for verifying the necessary requirements to render the content (encryption, scrambling, and others) and effectively renders the content for the end user.

On the server-side, is where a large part of the rights management responsibility lies. A set of decoupled components with a well-defined API that allows an integration between the necessary ones to implement the specific content business model. These services are the following:

- Content storage and distribution service: this service is responsible for the storage and distribution of user generated content in a protected manner;

- Content protection service: the service is responsible for the protection of the content. The content is protected by specific protection tools and specific protection mechanisms that may change according to the content and the business model that is going to be implemented;

- Content registration service: this service is responsible for registering the content on the platform that will be used to uniquely identify the content on the system. This unique identifier is used to identify the user generated content throughout the entire content lifecycle;

- Payment service: if the business model includes the possibility to trade content, this payment service is responsible to communicate with a payment gateway that implements the necessary mechanisms to process payments;

- Protection tools service: this service is responsible for the registration of content protection tools on the system and for making those tools available for the content protection service to use when implementing the content protection schemas (such as encryption, scrambling, watermarking and others);

- Authentication service: handles the registration of users and services on the system as well as the requests for authenticate users on behalf of other services;
- Licensing service: this is one of the most important services of the rights management framework, responsible for creating license templates, define and produce new content licenses (that represent the type of rights, permissions and/or restrictions of a given user, or group of users, over the content) and provide licenses, upon request, to specific users.

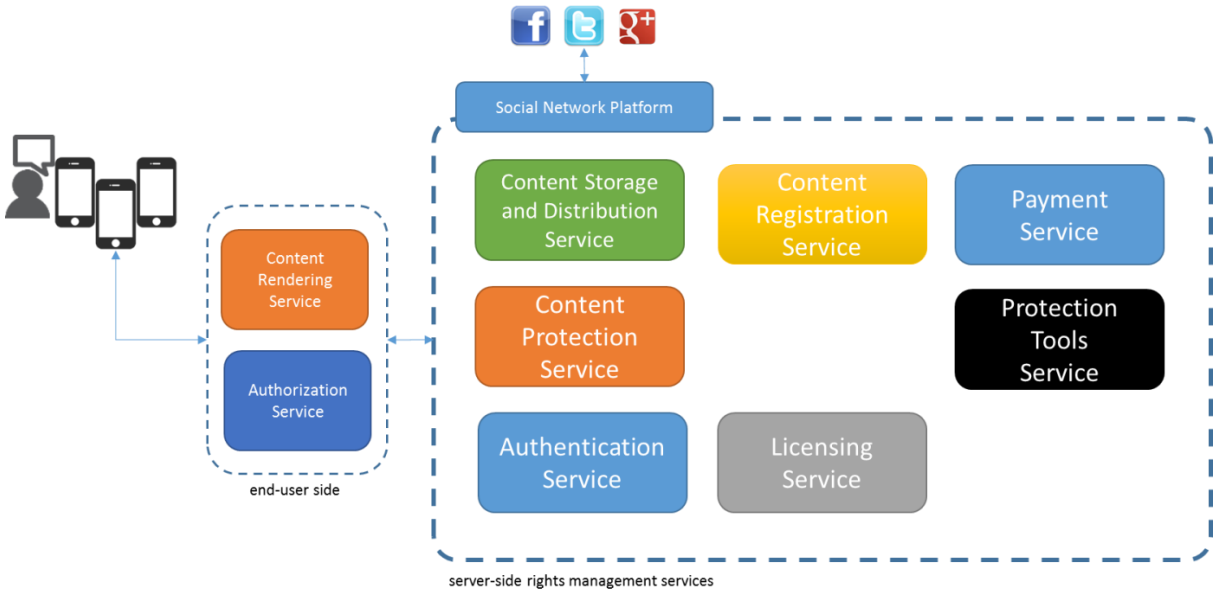

Figure 4. Overview of the architecture integrated with the rights management system

The following sections of this document will provide a description on how the user can utilize this platform to share user generated content on the social network, and how a user can access content shared by other users.

## Registration on the platform

This novel platform presupposes that all the system services are initially registered on that platform. This means that each one of the different services, either server-side or client-side have to be individually registered at the platform. This registration process assigns unique credentials to each one of the services, ensuring that they are uniquely registered and that these credentials will be used to identify and differentiate the services in future interactions (Figure 5). This registration process is conducted by the authentication service that on its turn issues credentials to all the other services and acts as a central trustworthy mechanism. Moreover, all the communication between the different services is conducted over a secure and authenticated channel, using Secure Sockets Layer/Transport Layer Security(SSL/TLS) – this ensures the authentication and security of the servers where the services are deployed and allowing the establishment of secure communication channels (Stephen A Thomas, 2000).

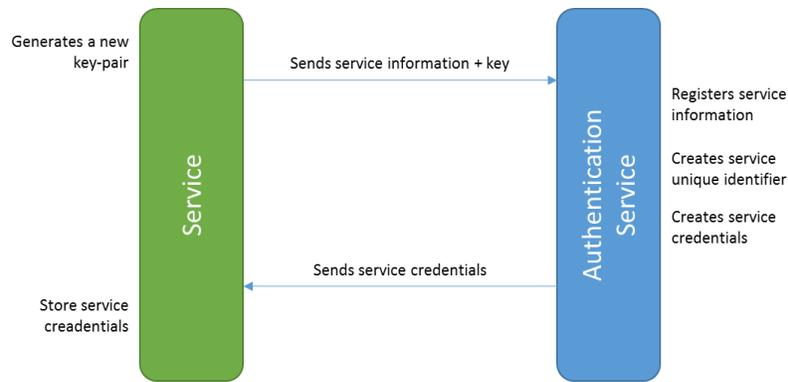

Figure 5 - Handling the registration of new services on the platform

1. The authentication service (AS) has cryptographic material ($K_{pub}^{AS}$, $K_{priv}^{AS}$) and credentials that were self-issued ($C^{AS}_{AS}$) or issued by other trustworthy entity ($C^{CA}_{AS}$);

2. The service that needs to be registered generates a key pair ($K_{pub}^{S}$, $K_{priv}^{S}$) and sends a registration request to the AS, passing some information about the service ($S_{info}$) and the public key ($K_{pub}^{S}$) of the service: $S_{info} + K_{pub}^{S}$;

3. AS receives this information, verifies it and then creates a unique service identifier ($S_{UUID}$). After this verification the AS creates the service credentials that will identify this service globally and uniquely on the platform: $C^{AS}_{S[UUID]} = K_{priv}^{AS}\{S_{UUID}, K_{pub}^{S[UUID]}, C^{AS}_{AS}\}^2$. These credentials, which are signed by AS, are then returned to the requesting service;

4. The requesting service, stores the credentials. This credential contains also the public key of the authentication service ($K_{pub}^{AS}$). This is used to prove this credentials to other entities that also rely on the same AS – services that trust AS, also trust on credentials issued by AS, presented by other services.

The service registration process, as described above needs to be repeated according to the number of services available within the social network platform. This enables the entire ecosystem of services to be trusted on that platform.

Another important aspect of the registration process concerns the registration of the users on the rights management platform. The registration of the user on the rights management platform can be dependent or independent of the social network platform. In the example that is presented here, it is assumed that this registration process is performed fully integrated with the social network platform.

This process performs in the following manner (Figure 6):

1. Assuming that the user still has no account on a social network platform, the user starts the registration process on the social network. In order to do that the user needs to supply its email address (as username) and a password;

2. The registration process on the social network platform finishes and a confirmation message is sent to the end-user;

3. Next, the user, using the client-side rights management authorization service (AUTS), initiates the registration process in the rights management platform. The AUTS presents several registration options to the end-user (integrated with some social network platforms -using either

---

[2] Some notes about the notation used: key(content) means the "content" is encrypted using "key"; key{content} represents "content" is signed using "key"; algo[content] means that "content" is hashed with the "algo" algorithm.

Oauth- or an independent mode). For this case, the user will use registration options by using the mode of integrated authentication;

4. The user introduces the social account credentials (email, password) on the AUTS that starts the authentication process on the social network platform. If successful, the social network returns an access token that has a specific validity and a set of permissions to conduct different operations on the social network on behalf of the user;

5. AUTS, using the user credentials (email, password) creates a secret key that is used to initialize a secure storage on the authorization service: $S_k^{SStorage}$ = SHA1[email+password];

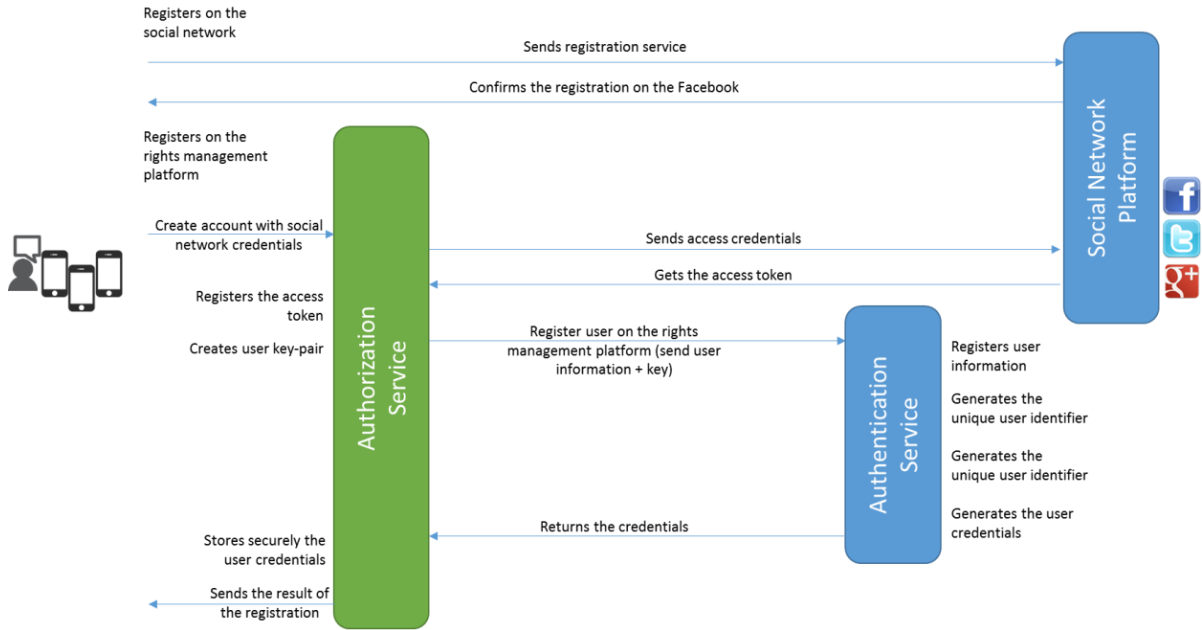

Figure 6 - Overview of the user registration process

6. The AUTS securely stores the user information, and the social network access token. Additionally, the AUTS creates a key-pair for the user ($K_{pub}^U$, $K_{priv}^U$) also storing it in a secure manner: $S_k^{SStorage}(K_{pub}^U, K_{priv}^U,$ user_info, token);

7. AUTS contacts the AS to register the user on the platform. This is performed using the $C^{AS}_{S[AUTS]}$ that contains the $K_{pub}^{AS}$. $C^{AS}_{S[AUTS]}$ is also sent to ensure that the AUTS has been previously registered: $K_{pub}^{AS}$ (email, $K_{pub}^U, C^{AS}_{S[AUTS]}$);

8. The AUTS receives all this information and after deciphering it, and validating the AUTS credential, registers the user information, generates a unique identifier for the user and creates credentials for the user: $C^{AS}_{UUID} = K_{priv}^{AS} \{UUID, K_{pub}^U\}$;

9. The credentials are returned to the AUTS and are securely stored: $S_k^{SStorage}(C^{AS}_{UUID})$. The user is notified about the result of the registration operation.

This is the concluding step of the service and user registration on the rights management platform. The user is now prepared to use both the rights management service and the social network platform.

## Sharing content on the platform

The other important functionality on the system is the sharing of user generated content (UGC) on the social network. This sharing mechanism is performed through the rights management platform, and the content is stored securely on a configured location (it can be on a specific storage location, on the social platform or on the rights management platform). When the user uploads user generated content, the content is protected and the rights, permissions and restrictions about the content can be defined by the user.

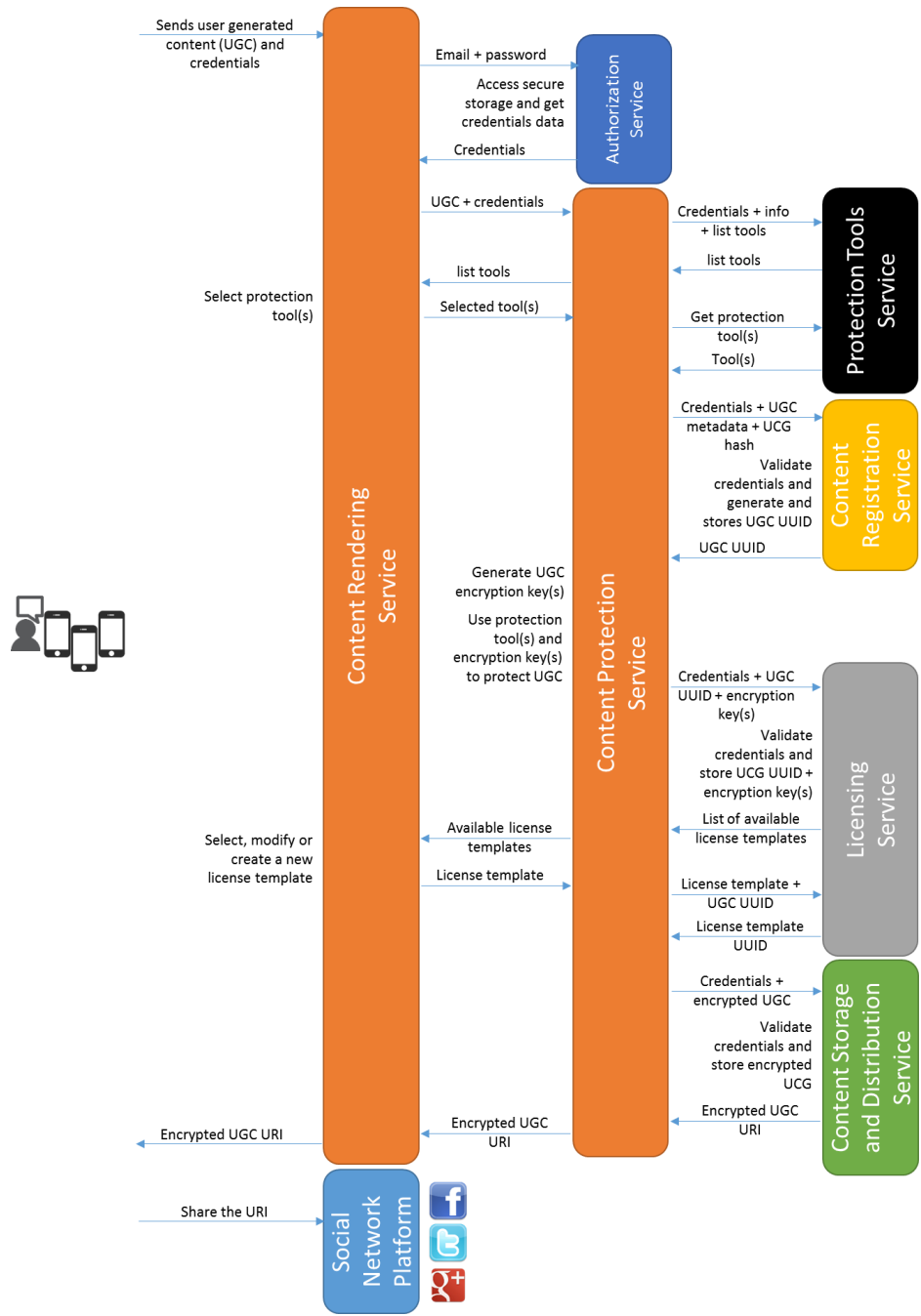

Figure 7 - Overview of the user generated content secure sharing on the social network platform

This process assumes that both user that generates the content and the users willing to access the content are properly registered and authenticated on the social network platform and on the rights management platform (Figure 7).

In a brief discussing way, the user generated content is uploaded to the rights management platform, the access rights and permissions are defined by the user, the content is protected, and a URI is returned to be shared on the social network platform.

The novel content sharing process, using the mechanisms described in this chapter, can be now defined in the following steps:

1. The user sends the user generated content (UGC) that it expects to share on the social network. This UGC is uploaded through the content rendering service (CRS). This service requires the user to enter its credentials (email and password), if the user is not yet authenticated. These credentials are used to access the secure storage: $S_k^{SStorage}$ = SHA1[email+password];

2. The CRS contacts the AUTS, which reads from the secure storage the user rights management system credentials: $C^{AS}_{UUID}$;

3. The CRS uploads to the content protection service (CPS) the UGC and sends the user credentials, obtained in the previous step: $UGC_{UUID}$, $C^{AS}_{UUID}$;

4. The CPS, after retrieving some metadata information about the UGC (such as the type, the format, the encoding, among others), contacts the protection tools service (PTS), requesting a list of available protection tools, that can be suitable to protect the UGC. The PTS sends its credentials and some information about the content: $C^{AS}_{CPS}$, UGC_info;

5. The PTS also returns a list of protection tools that match the request made by the CPS. This information is signed by PTS: $K_{priv}^{PTS}$\{protection_tools_list\};

6. The CPS returns the list of protection tools to the CRS, and presents it to the user. The user selects the most appropriate protection tools, adjusting the parameters of applicability of the tools to the UGC and submits its request about the necessary protection tools;

7. The CPS requests the selected protection tools from the protection tools service. The PTS returns the requested tools to the CPS;

8. Next, the CPS requests to the content registration service for the UGC to be registered. For this, the CPS send its credentials, the UCG metadata and the content hash: $C^{AS}_{CPS}$, UGC_info, SHA1[UGC];

9. The content registration service (CRGS), stores the received information, and generates a unique content identifier that is returned to the content protection service: $K_{priv}^{CRS.}$\{$UGC_{UUID}$\};

10. The CPS generates one or more content encryption keys (CEK[1], CEK[2] … CEK[n]) that are applied over the UGC, using the selected protection tools, in order to ensure the appropriate content protection;

11. Following this protection process, the CPS sends the content encryption keys for registration at the licensing service. Each of the content encryption keys is protected with the user key, and the entire message is protected by the CPS key: $C^{AS}_{CPS}$, $K_{pub}^{CPS}(K_{pub}^{U}$(CEK[1], CEK[2] … CEK[n]), $UGC_{UUID}$);

12. The licensing service (LS) after validating all the received information, returns a list of licensing templates to the content protection service. The CPS returns the list of licensing templates to CRS, and the user can select the most appropriate license template, modify it and adapt it, or simply create a new one;

13. The license template ($LIC_{TPL}$) is sent to the CPS that after sends it to the licensing service and associates it with the identifier of the UGC: $LIC_{TPL}$, $UGC_{UUID}$. The licensing service returns the license template identifier ($LIC_{TPL}$ [UUID]);

14. In the next stage, the CPS sends the protected UGC to the content storage and distribution service that stores the encrypted content: $C^{AS}_{CPS}, K_{priv}^{CPS}\{CEK[n](UGC), UGC_{UUID}\}$;

15. The content storage and distribution service returns a URI for the location of the stored encrypted UGC. This URI is returned to the user that can share it on the social network platform afterwards.

After this process is completed, the UGC shared by the user is shared on the social network platform. The user can also use the social network sharing mechanisms as a way to control how the UGC is propagated on the social network. But, in order to have a fine grained control over the UGC, the user needs to use the rights management system to produce specific licenses with the conditions under which the UGC can be used. These licenses are produced in multiple formats (either in ODRL or MPEG-21 REL). In addition, these licenses are used to support the expression of rights over the UGC. Therefore, when the user uploads user generated content to the rights management system, and after the process that was described previously, the subsequent steps are the following:

1. The CPS contacts licensing service to obtain the appropriate license template for the specific UGC, which was previously created: $LIC_{TPL}[UUID]$. The license template is an XML-formatted document that contains parameterized fields that can be adapted to specific rights situations;

2. A typical license template for user generated content would be composed by following elements:
    a. User unique identifier (UUID), multiple users ($UUID_1, UUID_2,…, UUID_n$) or a group identifier ($G_{UUID}$): these fields represent the unique identifiers of the users or groups to whom the user generated content is going to be shared;
    b. The unique identifier of the content: $UGC_{UUID}$;
    c. List of permissions ($Permission_1…Permission_n$);
    d. List of restrictions ($Restriction_1…Restriction_n$);
    e. Validity date (validity);
    f. The different content encryption keys ($CEK[1], CEK[2] … CEK[n]$). The content encryption keys are protected with user public key: $K_{pub}^{U}(CEK[1], CEK[2] … CEK[n])$;
    g. The license signature, where the license contents are signed by the licensing service: License = $K_{priv}^{LIS}\{UUID_1.. UUID_n, G_{UUID}1, G_{UUID}n, UGC_{UUID}, Permission_1 ..Permission_n, Restriction_1 ..Restriction_n, Validity, K_{pub}^{U}(CEK[1] … CEK[n])\}$.

3. The license is stored on the licensing service, where it can be accessed by legitimate users.

### Accessing content on the platform

Finally the last process in this case-study is to present how the users can access user generated content that was shared by other users on the social network platform. In order to do that, the user needs to be registered on the social network platform and on the rights management system.

When navigating through the timeline of the social network platform, user generated content that was shared over the social network platform, is presented in the form of a special URI, that, when clicked, is intercepted by the rights management platform, and the access process is started.

The referred process is described in the following steps:

1. The CRS, while trying to render the content that is shared on the social network platform, detects that it is protected content, and contacts the authorization service to access the appropriate information to try rendering the content;

2. The user authenticates to the system using the authorization service, supplying its credentials (email and password) to unlock the secure storage and retrieve the user information;

3. The authorization service, using the $UGC_{UUID}$ embedded on the URI, checks if a license for this UGC already exists on the secure storage. If a license already exists:

    a. The authorization service checks the license contents, validating the license digital signature and verifying the $UGC_{UUID}$;

    b. If the $UGC_{UUID}$ is the right one, the Validity is checked and the list of permissions and restrictions are evaluated;

    c. If the conditions are met, the content can be deciphered and rendered by the content rendering service. The content encryption keys can be retrieved from the license, and used to decipher the content: $K_{priv}^{U}(K_{pub}^{U}(CEK[1] \ldots CEK[n])) = CEK[1] \ldots CEK[n]$;

    d. Content is rendered by the CRS while the license conditions are fulfilled;

4. If the authorization still does not possess a valid license for the $UGC_{UUID}$ that the content rendering service is trying to render to view, the following process should occur:

    a. The user authenticates to the system using the authorization service, supplying its credentials (email and password) to unlock the secure storage and retrieve the user information;

    b. The authorization service, after getting the appropriated user information, including the credentials, from the secure storage, allows the CRS to contact the licensing service, passing its credentials ($C^{AS}_{CRS}$), the user credentials ($C^{AS}_{UUID}$) and the user generated content identifier ($UGC_{UUID}$) the user is trying to render;

    c. The licensing service receives and validates the data that was sent by the CRS, and uses the user generated content unique identifier ($UGC_{UUID}$) and the user unique identifier (UUID) to verify the existence of a valid license. If the license exists on the system, that license is returned to the CRS, that passes it, for validation and storage, to the authorization service: License = $K_{priv}^{LIS}$ {$UUID_1$.. $UUID_n$ , $G_{UUID}1$ , $G_{UUID}n$ , $UGC_{UUID}$, $Permission_1$ ..$Permission_n$ , $Restriction_1$ ..$Restriction_n$, Validity, $K_{pub}^{U}$ ($CEK[1] \ldots CEK[n]$)};

    d. The authorization service validates the license signature, verifying its contents and validity and asserting the correct $UGC_{UUID}$;

    e. If the $UGC_{UUID}$ is the right one, the Validity is checked and the list of permissions and restrictions are evaluated;

    f. If the conditions are met, the content can be deciphered and rendered by the CRS. The content encryption keys can be retrieved from the license, and used to decipher the content: $K_{priv}^{U}(K_{pub}^{U}(CEK[1] \ldots CEK[n])) = CEK[1] \ldots CEK[n]$;

    g. Content is rendered by the CRS while the license conditions are satisfied.

After this process is executed, the access to the CRS can be granted or not, depending on the conditions expressed on the license. For simplicity sake, there are several other processes that were not included in this description, such as, for instance, the verification of the protection mechanisms that were applied to the content, and the download of the appropriated mechanisms to allow the local temporarily unprotected version of the user generated content to be rendered.

The usage of rights management systems to offer security and privacy to shared user generated content, offers additional privacy and security mechanisms that are out of the control of the social network platform itself (Rodríguez, Rodríguez, Carreras, & Delgado, 2009). The users can take advantage of both (the rights management system and the social network platform) to offer a finer control on the content sharing privacy and security properties. This is a novel approach (Marques & Serrão, 2013b) that clearly puts the security and privacy control on the end-user side.

## A CASE STUDY FOR AN INTELLIGENT INTRUSION DETECTION/PREVENTION SYSTEM ON A SOFTWARE-DEFINED NETWORK

This section presents and discusses a case study about an intelligent Intrusion Detection/Prevention System (IDS/IPS) belonging to a software-defined network. In this case study, the IDS/IPS behavior is controlled by a Kinetic module (Feamster, 2014). The Kinetic language (Monsanto, Reich, Foster, Rexford, & Walker, 2013) is an SDN control framework where operators can define a network policy as a Finite State Machine (FSM). The transitions between states of a FSM can be triggered by different types of dynamic events in the network, (e.g. intrusion detection, host state). Based on different network events, operators can enforce different policies to the network using an intuitive FSM model. Kinetic is implemented as a Pyretic controller module written in Python. In order to acquire more details related to Pyretic and Python, consult respectively (Pyretic, 2014) and (Python, 2014).

"A Kinetic control program permits programmer-defined events to dynamically change forwarding behavior for an arbitrary set of flows. Such events can range from topology changes (generated by the Pyretic runtime) to security incidents (generated by an intrusion detection system). The programmer specifies an FSM description that contains set of states, each of which maps to some network behavior that are encoded using Pyretic's policy language; and a set of transitions between those states, each of which may be triggered by events that the operator defines" (Feamster, 2014). In order to obtain further information on Kinetic, consult (Monsanto et al., 2013).

In this case study, an implementation of an IDS/IPS security module will be developed, which should behave as follows:

- If a host is infected and is not a privileged host then it is dropped;

- If a host is infected and is a privileged (exempt) host then the traffic from that host is automatically redirected to a garden wall host, where some corrective security actions could be issued over that infected host (e.g. clean and install security patches for trying to recover it);

- If a host is not infected then the traffic from that host is forwarded towards its final destination.

### Code Explanation

In Table 1, it is displayed a partial view of the Python code that implements the Kinetic control program that will be used in this section to evaluate the intelligent IDS/IPS. To become clearer, this code functionality is explained in the following paragraph.

Each time a new packet arrives to the system, the IDS/IPS initially processes that packet and defines the policy to be applied to that packet (i.e. drop | redirect | forward). This policy is then delivered to a second module that implements further MAC functionality, namely the learning algorithm of MAC addresses to enhance the L2 packet forwarding. This second module is the one that effectively forwards or redirects the packet (otherwise if the packet is to be drooped, this second module will not receive any packet at all because it was already discarded by the first IDS/IPS module).

The code shown in Table 1 corresponds to the IDS/IPS module and has its code encapsulated inside a class designated by "gardenwall", which was instantiated from class "DynamicPolicy" (to support the processing of JSON events, as it will be explained below). The function "lpec" is like a packet input filter because it only selects the packets whose source IP address is specified by variable *srcip*. This aims to process the first packet of a flow exactly in the same way as all the following packets of that flow. In this example, a transition function encodes logic that indicates the new value a state variable should take when a particular event arrives at the controller. For example, the *infected* transition function encodes a single case: when an *infected* event occurs, the new value taken by the state variable *infected* is the value of that event (i.e. FALSE or TRUE). This is an example of an exogenous transition (i.e. the state variable *infected* is changed by an external event); other exogenous transition in this scenario is the one associated with the state variable *exempt*. In opposition, the transition associated to the state variable *policy* is endogenous because its state is triggered by both internal state variables of the current FSM: *infected* and *exempt*.

Table 1 - Partial view of the code for the module IDS/IPS.

```
class gardenwall(DynamicPolicy):
    def __init__(self):
        # Garden Wall
        def redirectToGardenWall():
            client_ips = [IP('10.0.0.1'), IP('10.0.0.2')]
            rewrite_policy = rewriteDstIPAndMAC(client_ips, '10.0.0.3')
            return rewrite_policy
        ### DEFINE THE LPEC FUNCTION
        def lpec(f):
            return match(srcip=f['srcip'])
        ## SET UP TRANSITION FUNCTIONS
        @transition
        def exempt(self):
            self.case(occurred(self.event),self.event)
        @transition
        def infected(self):
            self.case(occurred(self.event),self.event)
        @transition
        def policy(self):
            # If exempt, redirect pkt to gardenwall;rewrite dstip to 10.0.0.3
            self.case(test_and_true(V('exempt'),V('infected')),
C(redirectToGardenWall()))
            # If infected, drop pkt
            self.case(is_true(V('infected')),C(drop))
            # Else, identity -> forward pkt
            self.default(C(identity))
         ### SET UP THE FSM DESCRIPTION
        self.fsm_def = FSMDef(
            infected=FSMVar(type=BoolType(),
                            init=False,
                            trans=infected),
```

```
                exempt=FSMVar(type=BoolType(),
                        init=False,
                        trans=exempt),
            policy=FSMVar(type=Type(Policy,
                    {drop,identity,redirectToGardenWall()}),
                    init=identity,
                    trans=policy))
        ### SET UP POLICY AND EVENT STREAMS
        fsm_pol = FSMPolicy(lpec,self.fsm_def)
        json_event = JSONEvent()
        json_event.register_callback(fsm_pol.event_handler)
        super(gardenwall,self).__init__(fsm_pol)
```

The Finite State Machine (FSM) (see Figure 8) used in the current scenario associates the transition functions previously defined with the appropriate state variables. The FSM definition consists of a set of state variable definitions. Each variable definition simply specifies the variable's type (i.e., set of allowable values), initial value, and associated transition functions. The infected variable is a boolean whose initial value is FALSE (representing the assumption that hosts are initially not infected), and transitions based on the infected function defined previously. Likewise, the policy variable can take the values *drop* or *identity*, initially starts in the *identity* state, and transitions based on the policy function defined previously. The FSMPolicy that Kinetic provides automatically directs each incoming external event to the appropriate *lpec* FSM, where it will be handled by the exogenous transition function specified in the FSM description (i.e. the function *self.fsm_def*). In this way, it is ensured that the FSM works as expected.

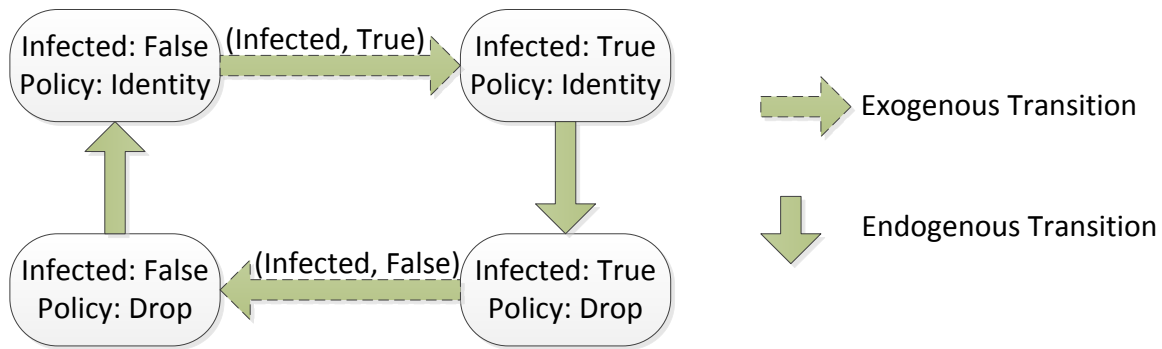

Figure 8– Finite State Machine (FSM) (Feamster, 2014)

## Evaluation

The network topology used in the current evaluation made with a network emulator is shown in Figure 9. All the evaluation was performed in a single Linux virtual machine (Ubuntu).

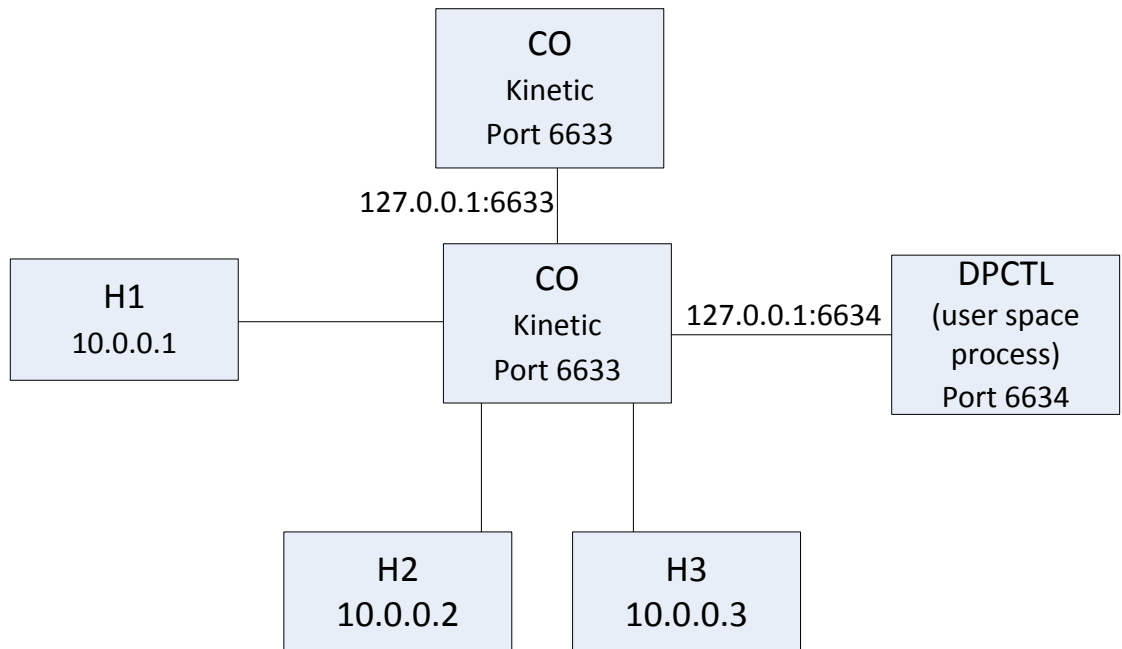

Figure 9 – Network Topology under test (Feamster, 2014)

We now initiate the evaluation, opening a Linux shell, and run our Kinetic controller application with the following commands:

$ cd ~/pyretic

$ pyretic.py pyretic.kinetic.examples.gardenwall

As shown in Figure 10, the kinetic controller prints out some results from a verification of network policies using the NuSMV symbolic model checker (NuSMV, 2014). Kinetic automatically generates a NuSMV input from the program written by the programmer/operator, and verifies logic statements written in CTL (Computation Tree Logic) (CTL, 2014).

```
*** Please report bugs to <numsv-users@fbk.eu>
*** Copyright (c) 2010, Fondazione Bruno Kessler
*** This version of NuMSV is linked to the CUDD library version 2.4.1
*** Copyright (c) 1995-2004, Regents of the University of Colorado
*** This version  of NuMSV is linked to the MiniSat  SAT solver.
*** See http://www.cs.chalmers.se/Cs/Research/FormalMethods/MiniSat
*** Copyright (c) 2003-2005, Niklas Een, Niklas Sorensson

-- specification AG ((infected & !exempt) -> AX policy = policy_1)   is true
-- specification AG (!infected -> AX policy = policy_2) is true
-- specification AG ((infected & exempt) -> AX policy = policy_3)    is true
-- specification A [ policy = policy_2  U infected ]    is true
system diameter: 3
reachable states: 12 (2^3.58496) out of 12 (2^3.58496)
====================== NuSMV Output  End ====================

POX 0.1.0 (betta)  / Copyright 2011-2013 James McCauley, et al.
Connected to pyretic frontend.
INFO:core:POX 0.1.0 (betta) is up.
```

Figure 10 – Kinetic controller terminal

In a second shell, we start the network emulator "mininet", performing the following command:

```
$ sudo mn --controller=remote --topo=single,3 --mac –arp
```

The obtained output result is visualized in Figure 11. In addition, the bottom line of Figure 12 shows that the Kinetic controller discovered the emulated switch.

```
mininet@mininet-vm:~$ sudo mn --controller=remote --topo=single, 3 --mac --arp
*** Creating network
*** Adding controller
*** Adding hosts:
h1 h2 h3
*** Adding switches:
s1
*** Adding links:
(h1, s1) (h2, s1) (h3, s1)
*** Configuring hosts
h1 h2 h3
*** Startting controller
*** Startting 1 switches
s1
*** Startting CLI:
mininet>
```

Figure 11 – Mininet terminal

```
*** Copyright (c) 2010, Fondazione Bruno Kessler
*** This version of NuMSV is linked to the CUDD library version 2.4.1
*** Copyright (c) 1995-2004, Regents of the University of Colorado
*** This version of NuMSV is linked to the MiniSat SAT solver.
*** See http://www.cs.chalmers.se/Cs/Research/FormalMethods/MiniSat
*** Copyright (c) 2003-2005, Niklas Een, Niklas Sorensson

-- specification AG ((infected & !exempt) -> AX policy = policy_1)   is true
-- specification AG (!infected -> AX policy = policy_2) is true
-- specification AG ((infected & exempt) -> AX policy = policy_3)   is true
-- specification A [ policy = policy_2  U infected ]    is true
system diameter: 3
reachable states: 12 (2^3.58496) out of 12 (2^3.58496)
====================== NuSMV Output  End =====================

POX 0.1.0 (betta)  / Copyright 2011-2013 James McCauley, et al.
Connected to pyretic frontend.
INFO:core:POX 0.1.0 (betta) is up.
INFO: openflow.of_01: [None 1] closed
INFO: openflow.of_01: [00-00-00-00-00-01 2] connected
```

Figure 12 – Switch contacts the controller

Imagine now the situation that host "h1" become compromised (infected). This situation originates the transmission of an event to change the state of the FSM in a way that any traffic originated in that host should be discarded in the switch. In this network status, the traffic ICMP between hosts "h1" and "h2" becomes blocked. We issue the transmission of the "infected" event to the controller executing in a third terminal the following command:

```
$ python json_sender.py -n infected -l True --flow="{srcip=10.0.0.1}" -a 127.0.0.1 -p 50001
```

The "infected" event was sent to the controller as it is possible to be visualized in Figure 13.

```
mininet@mininet-vm:~/pyretic/pyretic/kinetic$ python json_sender.py -n infected -l
True --flow ="{srcip=10.0.0.1}" -a 127.0.0.1 -p 50001

Flow_Str = {srcip=10.0.0.1}

Data Payload = {'dstip': None, 'protocol': None, 'srcmac': None, 'tos': None,
'vlan_pcp': None, 'dstmac': None, 'inport': None, 'switch': None, 'ethtype': None,
'srcip': '10.0.0.1', 'dstport': None, 'srcport': None, 'vlan_id': None}

Ok
mininet@mininet-vm:~/pyretic/pyretic/kinetic$
```

Figure 13 – JSON event transmitted to the controller

After some milliseconds, the Kinetic controller received the event informing that host h1 is infected (see Figure 14). As a consequence of this, the controller changed the policy to drop the packets originated by host "h1".

```
-- specification A [ policy = policy_2  U infected ]    is true
system diameter: 3
reachable states: 12 (2^3.58496) out of 12 (2^3.58496)
====================== NuSMV Output  End =====================

POX 0.1.0 (betta)  / Copyright 2011-2013 James McCauley, et al.
Connected to pyretic frontend.
INFO:core:POX 0.1.0 (betta) is up.
INFO: openflow.of_01: [None 1] closed
INFO: openflow.of_01: [00-00-00-00-00-01 2] connected
Received connection from ('127.0.0.1', 42143)
Received event infected is True related with flow {'srcip': 10.0.0.1}
fsm_policy:event_name= infected
fsm_policy:event_value= True
fsm_policy:event_state= {'policy': drop, 'infected': True, 'exempt': False}
fsm_policy:self.policy = if
   match:  ('srcip', 10.0.0.1)
then
   [DynamicPolicy]
   drop
else
   identity
```

Figure 14 – Controller changes the policy to drop

After this, we have tried to send two ping messages from host "h1" to host "h2" but as it is shown in Figure 15 without any success. This occurs because the IDS/IPS installed in the switch between "h1" and "h2" a policy to drop the packets originated by host "h1".

```
s1
*** Adding links:
(h1, s1) (h2, s1) (h3, s1)
*** Configuring hosts
h1 h2 h3
*** Startting controller
*** Startting 1 switches
s1
*** Startting CLI:
mininet> h1 ping -c 2 h2
PING 10.0.0.2 (10.0.0.2) 56(84) bytes of data.

--- 10.0.0.2 ping statistics ---
2 packets transmitted, 0 received, 100% packet loss, time 1008ms

mininet>
```

Figure 15 – ICMP traffic is dropped

Next, assuming that host "h1" was classified as a privileged (exempt) terminal, then the controller will be notified from this through the following event:

```
$ python json_sender.py -n exempt -l True --flow="{srcip=10.0.0.1}" -a 127.0.0.1 -p 50001
```

Almost immediately, the Kinetic controller received the event informing that host "h1" is infected (see Figure 16). As a consequence of this, the controller changed the policy to redirect the packets originated by host "h1" to host "h3" (policy modify) for further analysis. This policy is installed in the switch.

```
fsm_policy:event_name= infected
fsm_policy:event_value= True
fsm_policy:event_state= {'policy': drop, 'infected': True, 'exempt': False}
fsm_policy:self.policy = if
   match: ('srcip', 10.0.0.1)
then
   [DynamicPolicy]
   drop
else
   identity
Received connection from ('127.0.0.1', 42144)
Received event exempt is True related with flow {'srcip': 10.0.0.1}
fsm_policy:event_name= exempt
fsm_policy:event_value= True
fsm_policy:event_state= {'policy': modify: ('dstip', 10.0.0.3) ('dstmac',
00:00:00:00:00:03), 'infected':True, 'exempt': True}
fsm_policy:self.policy = if
   match: ('srcip', 10.0.0.1)
then
   [DynamicPolicy]
   modify: ('dstip', 10.0.0.3) ('dstmac', 00:00:00:00:00:03)
else
   identity
```

Figure 16 – Policy is changed to redirect the traffic

The redirection of traffic from host "h1", it is perfectly visible in Figure 17, after we repeat the ping command. One can note that the host replying to the ping is host "h3" instead of "h2". As already explained, host "h3" is responsible to recover in terms of security any privileged hosts that by some reason become compromised.

```
mininet> h1 ping -c 2 h2
PING 10.0.0.2 (10.0.0.2) 56(84) bytes of data.
64 bytes from 10.0.0.3 icmp_req=1 ttl=64 time=115 ms
64 bytes from 10.0.0.3 icmp_req=2 ttl=64 time=112 ms

--- 10.0.0.2 ping statistics ---
2 packets transmitted, 2 received, 0% packet loss, time 1002ms
rtt min/avg/mdev = 112.492/114.106/115.720/1.614 ms
mininet>
```

Figure 17 – ICMP traffic is redirected

After, some corrective actions performed in host "h1" by "h3", one can assume that host "h1" has recovered. In this way, a new event is sent to the controller notifying host "h1" changed to the state of "not infected", as follows:

```
$ python json_sender.py -n infected -l False --flow="{srcip=10.0.0.1}" -a 127.0.0.1 -p 50001
```

Figure 18 illustrates some controller´s output informing that the last event was received and the forwarding policy changed to forward the traffic towards host "h2" (policy identity).

```
match: ('switch', 1) ('dstmac', 00:00:00:00:00:01)
then
   [DynamicPolicy]
   fwd 1
else
   flood on:
   -------------------------------------------------------------------------------------------------
   switch  |   switch edges   |   egress ports                                                     |
   -------------------------------------------------------------------------------------------------
   1       |                  | 1[2]---, 1[3]---, 1[1]---                                          |
Received connection from ('127.0.0.1', 42145)
Received event infect is False related with flow {'srcip': 10.0.0.1}
fsm_policy:event_name= infected
fsm_policy:event_value= False
fsm_policy:event_state= {'policy': identity, 'infected': False, 'exempt': True}
fsm_policy:self.policy = if
   match: ('srcip', 10.0.0.1)
then
   [DynamicPolicy]
   identity
else
   identity
```

Figure 18 – Policy returns back to traffic pass through (identity)

From Figure 19 is possible to conclude that host "h1" is now receiving response from host "h2" itself.

```
        64 bytes from 10.0.0.3 icmp_req=1 ttl=64 time=115 ms
        64 bytes from 10.0.0.3 icmp_req=2 ttl=64 time=112 ms

        --- 10.0.0.2 ping statistics ---
        2 packets transmitted, 2 received, 0% packet loss, time 1002ms
        rtt min/avg/mdev = 112.492/114.106/115.720/1.614 ms
        mininet> h1 ping -c 2 h2
        PING 10.0.0.2 (10.0.0.2) 56(84) bytes of data.
        64 bytes from 10.0.0.2 icmp_req=1 ttl=64 time=113 ms
        64 bytes from 10.0.0.2 icmp_req=2 ttl=64 time=59.5 ms

        --- 10.0.0.2 ping statistics ---
        2 packets transmitted, 2 received, 0% packet loss, time 1001ms
        rtt min/avg/mdev = 59.529/86.713/113.897/27.184 ms
        mininet>
```

Figure 19 – ICMP traffic is reaching again host h2

At this point, we finish our current evaluation of the intelligent IDS/IPS system. This is an important feature to be incorporated in Big Data scenarios once it facilitates the identification and solving of some attacks that a distributed Big Data architecture (in different phases of the Big Data lifecycle – from data capture to data processing and consumption) can suffer.

## BIG DATA SECURITY: FUTURE DIRECTIONS

Throughout this chapter it was possible to present some of the most important security and privacy challenges that affect Big Data projects and their specificities. Although the information security practices, methodologies and tools to ensure the security and privacy of the Big Data ecosystem already exist, the particular characteristics of Big Data make them ineffective if they are not used in an integrated manner. This chapter also presents some solutions for these challenges, but it does not provide a definitive solution for the problem. It rather points to some directions and technologies that might contribute to solve some of the most relevant and challenging Big Data security and privacy issues.

Next, two different use cases were presented. Both of the use-cases present some directions that contribute to solving part of the large Big Data security and privacy puzzle. In the first use-case it was presented an approach that tries solving security and privacy issues on social network user generated content. In this approach, an open an interoperable rights management system was proposed as a way to improve the privacy of users that share content over social networks. The processes described show how the rights management system puts the end-users on the control of their own user-generated content, and how they prevent abuses from either other users or the social network platform itself. The second use-case presented the capabilities offered by SDN in increasing the ability to collect statistics data from the network and of allowing controller applications to actively program the forwarding devices, are powerful for proactive and smart security policy enforcement techniques such as active security (Hand, Ton, & Keller, 2013). This novel security methodology proposes a novel feedback loop to improve the control of defense mechanisms of a networked infrastructure, and is centered around five core capabilities: protect, sense, adjust, collect, counter (Kreutz et al., 2014). In this perspective, active security provides a centralized programming interface that simplifies the integration of mechanisms for detecting attacks, by i) collecting data from diverse sources (to identify attacks with more assertiveness), ii) converging to a consistent policy configuration for the security appliances, and iii) enforcing countermeasures to block or minimize the effect of such attacks. Previous aspects were partially covered by our IDS/IPS case study but notably need to be further developed and are an important contribution to the security and privacy of Big Data ecosystem.

As noted throughout this chapter, although some important steps are being given towards solving Big Data security and privacy issues, there is still a long road ahead. In the conclusion of this chapter, the authors would like to refer some interesting topics where the research community could work actively to develop new Big Data security and privacy solutions.

Research challenges in this Big Data ecosystem range from the data creation (and the Big Data sources - devices), data storage and transportation, data transformation and processing, and finally data usage. To support this lifecycle, a high capacity and highly distributed architecture will be necessary, exposed to an hostile environment subject to all kinds of attacks. The SDN approach as proposed on this chapter is a possible solution to counter these threats, however further research needs to be conducted, in particular on what concerns to automatic adaptation of switching and behavior-based security policies (P. Chen, Jorgen, & Yuan, 2011; Dohi & Uemura, 2012).

There are also important research challenges on maintaining end to end data security and privacy. Ensuring that data is never revealed in clear, in particular to non-authorized parties, on any point of the Big Data lifecycle. Moving from data to programs, there are techniques for protecting privacy in browsing, searching, social interactions, and general usage through obfuscation methods. However, there is more research to be conducted on the processing of encrypted data and privacy protection in the context of both computer programs and web-based systems.

More research challenges in the Big Data area include developing techniques to perform a transparent computations over encrypted data with multiple keys, from multiple sources and multiple users. In terms of research it would be challenging to study and develop ways to delegate limited functions over encrypted data, so that third parties can analyze it. All the aspects related with key management, authorization delegation, management of rights, are topics that require further research in this field.

When considering secure and private-aware system, trust is everything. In particular, in the case of Big Data, a trustworthy environment should be established for most of the scenarios (healthcare, assisted living, SCADA systems and many others). It is particularly challenging in terms of research directions how this environment can be attained. Trusting applications that are capable of querying and processing Big Data and extract knowledge from it and, trusting devices that collect all the data from multiple sources, constitute a basic security requirement. Understand how trust can be established among the end-users, the devices (IoT) and the applications is a hot research topic for the coming years.

On what concerns Big Data, these research challenges represent only the tip of the iceberg about the problems that still need to be studied and solved on the development of secure and privacy-aware Big Data ecosystem.

## REFERENCES

Advantech. (2013). *Enhancing Big Data Security*. Retrieved from http://www.advantech.com.tw/nc/newsletter/whitepaper/big_data/big_data.pdf

Agrawal, D., Das, S., & El Abbadi, A. (2011). Big data and cloud computing. In *Proceedings of the 14th International Conference on Extending Database Technology - EDBT/ICDT '11* (p. 530). New York, New York, USA: ACM Press. doi:10.1145/1951365.1951432

Carlos Serrão, Neves, D., Trevor Barker, & Massimo Balestri. (2003). OpenSDRM -- An Open and Secure Digital Rights Management Solution. In *Proceedings of the IADIS International Conference e-Society*.

Chen, P., Jorgen, B., & Yuan, Y. (2011). Software behavior based trusted attestation. In *Proceedings - 3rd International Conference on Measuring Technology and Mechatronics Automation, ICMTMA 2011* (Vol. 3, pp. 298–301). doi:10.1109/ICMTMA.2011.645

Chen, X., & Shi, S. (2009). A literature review of privacy research on social network sites. In *Multimedia Information Networking and Security, 2009. MINES'09. International Conference on* (Vol. 1, pp. 93–97).

Cloud Security Alliance. (2013). *Expanded Top Ten Security and Privacy Challenges*. Retrieved from https://downloads.cloudsecurityalliance.org/initiatives/bdwg/Expanded_Top_Ten_Big_Data_Security_and_Privacy_Challenges.pdf

CTL. (2014). *Computation tree logic*. Retrieved July 17, 2014, from http://en.wikipedia.org/wiki/Computation_tree_logic

DARPA. (2014). *MINING AND UNDERSTANDING SOFTWARE ENCLAVES (MUSE)*. Retrieved August 03, 2014, from http://www.darpa.mil/Our_Work/I2O/Programs/Mining_and_Understanding_Software_Enclaves_(MUSE).aspx

De Cristofaro, E., Soriente, C., Tsudik, G., & Williams, A. (2012). Hummingbird: Privacy at the time of twitter. In *Security and Privacy (SP), 2012 IEEE Symposium on* (pp. 285–299).

Demchenko, Y., Ngo, C., Laat, C. de, Membrey, P., & Gordijenko, D. (2014). Big Security for Big Data: Addressing Security Challenges for the Big Data Infrastructure. In W. Jonker & M. Petković (Eds.), *Secure Data Management* (pp. 76–94). Springer International Publishing. Retrieved from http://link.springer.com/chapter/10.1007/978-3-319-06811-4_13

Dohi, T., & Uemura, T. (2012). An adaptive mode control algorithm of a scalable intrusion tolerant architecture. In *Journal of Computer and System Sciences* (Vol. 78, pp. 1751–1754). doi:10.1016/j.jcss.2011.10.022

Feamster, N. (2014). *Software Defined Networking*. Retrieved August 02, 2014, from https://www.coursera.org/course/sdn

Gentry, C. (2009). *A fully homomorphic encryption scheme*. Stanford University. Retrieved from http://cs.au.dk/~stm/local-cache/gentry-thesis.pdf

Gentry, C. (2010). Computing arbitrary functions of encrypted data. *Communications of the ACM*. doi:10.1145/1666420.1666444

Goldwasser, S., Gordon, S. D., Goyal, V., Jain, A., Katz, J., Liu, F.-H., … Zhou, H.-S. (2014). Multi-input functional encryption. In *Advances in Cryptology--EUROCRYPT 2014* (pp. 578–602). Springer.

Google. (2014). *Encrypted Big Query Client*. Retrieved August 03, 2014, from https://code.google.com/p/encrypted-bigquery-client/

Gross, R., & Acquisti, A. (2005). Information revelation and privacy in online social networks. In *Proceedings of the 2005 ACM workshop on Privacy in the electronic society* (pp. 71–80).

Hand, R., Ton, M., & Keller, E. (2013). Active security. In *Proceedings of the Twelfth ACM Workshop on Hot Topics in Networks - HotNets-XII* (pp. 1–7). New York, New York, USA: ACM Press. doi:10.1145/2535771.2535794

Hasan, O., Habegger, B., Brunie, L., Bennani, N., & Damiani, E. (2013). A Discussion of Privacy Challenges in User Profiling with Big Data Techniques: The EEXCESS Use Case. In *2013 IEEE International Congress on Big Data* (pp. 25–30). IEEE. doi:10.1109/BigData.Congress.2013.13

HP. (2014). *Internet of Things Research Study* (p. 4). Retrieved from http://fortifyprotect.com/HP_IoT_Research_Study.pdf

*IBM big data platform - Bringing big data to the Enterprise*. (2014, July). CT000.

IDC. (2012). *Big Data in 2020*. Retrieved from http://www.emc.com/leadership/digital-universe/2012iview/big-data-2020.htm

Juels, A., & Oprea, A. (2013). New approaches to security and availability for cloud data. *Communications of the ACM*, *56*(2), 64. doi:10.1145/2408776.2408793

Jutla, D. N., Bodorik, P., & Ali, S. (2013). Engineering Privacy for Big Data Apps with the Unified Modeling Language. In *2013 IEEE International Congress on Big Data* (pp. 38–45). IEEE. doi:10.1109/BigData.Congress.2013.15

Kim, C., Jin, M.-H., Kim, J., & Shin, N. (2012). User perception of the quality, value, and utility of user-generated content. *Journal of Electronic Commerce Research*, *13*(4), 305–319.

Kindervag, J., Balaouras, S., Hill, B., & Mak, K. (2012). *Control And Protect Sensitive Information In the Era of Big Data*.

Kindervag, J., Wang, C., Balaouras, S., & Coit, L. (2011). *Applying Zero Trust To The Extending Enterprise*.

Kreutz, D., Ramos, F. M. V., Verissimo, P., Rothenberg, C. E., Azodolmolky, S., & Uhlig, S. (2014). Software-Defined Networking: A Comprehensive Survey, 49. Networking and Internet Architecture. Retrieved from http://arxiv.org/abs/1406.0440

Lindell, Y., & Pinkas, B. (2002). Privacy Preserving Data Mining. *Journal of Cryptology*, *15*(3), 177–206. doi:10.1007/s00145-001-0019-2

Luo, H., Lin, Y., Zhang, H., & Zukerman, M. (2013). Preventing DDoS attacks by identifier/locator separation. *IEEE Network*, *27*(6), 60–65. doi:10.1109/MNET.2013.6678928

Marques, J., & Serrão, C. (2013a). Improving Content Privacy on Social Networks Using Open Digital Rights Management Solutions. *Procedia Technology*, *9*, 405–410. doi:10.1016/j.protcy.2013.12.045

Marques, J., & Serrão, C. (2013b). Improving user content privacy on social networks using rights management systems. *Annals of Telecommunications - Annales Des Télécommunications*, *69*(1-2), 37–45. doi:10.1007/s12243-013-0388-1

McKenzie, P. J., Burkell, J., Wong, L., Whippey, C., Trosow, S. E., & McNally, M. B. (2012, June 6). User-generated online content: overview, current state and context. *First Monday*. Retrieved from http://firstmonday.org/ojs/index.php/fm/article/view/3912/3266

Michael, K., & Miller, K. W. (2013). Big Data: New Opportunities and New Challenges [Guest editors' introduction]. *Computer*, *46*(6), 22–24. doi:10.1109/MC.2013.196

MIT. (2014). *Big Data Privacy Workshop, Advancing the state of the art in Technology and Practice - Workshop summary report*. Retrieved from http://web.mit.edu/bigdata-priv/images/MITBigDataPrivacyWorkshop2014_final05142014.pdf

Monsanto, C., Reich, J., Foster, N., Rexford, J., & Walker, D. (2013). Composing software-defined networks. *Proceedings of the 10th USENIX Conference on Networked Systems Design and Implementation*, 1–14. Retrieved from http://dl.acm.org/citation.cfm?id=2482626.2482629\nhttp://www.frenetic-lang.org/pyretic/

NIST. (2014). *Framework for Improving Critical Infrastructure Cybersecurity*. Retrieved from http://www.nist.gov/cyberframework/upload/cybersecurity-framework-021214-final.pdf

NuSMV. (2014). *An overview of NuSMV*. Retrieved July 23, 2014, from http://nusmv.fbk.eu/NuSMV/

Okhravi, H., Hobson, T., Bigelow, D., & Streilein, W. (2014). Finding Focus in the Blur of Moving-Target Techniques. *IEEE Security & Privacy*, *12*(2), 16–26. doi:10.1109/MSP.2013.137

OWASP. (2014). *OWASP Internet of Things Top Ten Project*. Retrieved August 05, 2014, from https://www.owasp.org/index.php/OWASP_Internet_of_Things_Top_Ten_Project

Popa, R., & Redfield, C. (2011). Cryptdb: protecting confidentiality with encrypted query processing. *Proceedings of the …*, 85–100. doi:10.1145/2043556.2043566

Popa, R., & Redfield, C. (2012). CryptDB: Processing queries on an encrypted database. *Communications of the …*, *55*, 103. doi:10.1145/2330667.2330691

Pyretic. (2014). *Pyretic Language*. Retrieved August 05, 2014, from https://github.com/frenetic-lang/pyretic/wiki/Language-Basics

Python. (2014). *Python Language*. Retrieved August 03, 2014, from https://www.python.org/

Rivera, J., & van der Meulen, R. (2014). *Gartner Says the Internet of Things Will Transform the Data Center*. Retrieved August 05, 2014, from http://www.gartner.com/newsroom/id/2684915

Rodríguez, E., Rodríguez, V., Carreras, A., & Delgado, J. (2009). A Digital Rights Management approach to privacy in online social networks. In *Workshop on Privacy and Protection in Web-based Social Networks (within ICAIL'09), Barcelona*.

Serrão, C. (2008). *IDRM - Interoperable Digital Rights Management: Interoperability Mechanisms for Open Rights Management Platforms*. Universitat Politècnica de Catalunya. Retrieved from http://repositorio-iul.iscte.pt/handle/10071/1156

Serrão, C., Dias, J. M. S., & Kudumakis, P. (2005). From OPIMA to MPEG IPMP-X: A standard's history across R&D projects. *Signal Processing: Image Communication*, *20*(9), 972–994.

Serrão, C., Dias, M., & Delgado, J. (2005). Using Web-Services to Manage and Control Access to Multimedia Content. *ISWS05-The 2005 International Symposium on Web Services and Applications, Las Vegas, USA*.

Serrão, C., Rodriguez, E., & Delgado, J. (2011). Approaching the rights management interoperability problem using intelligent brokerage mechanisms. *Computer Communications*, *34*(2), 129–139.

Stephen A Thomas. (2000). *SSL & TLS Essentials: Securing the Web* (Pap/Cdr., p. 224). Wiley.

Tankard, C. (2012). Big data security. *Network Security*, *2012*(7), 5–8. doi:10.1016/S1353-4858(12)70063-6

Van Dijk, M., Gentry, C., Halevi, S., & Vaikuntanathan, V. (2010). Fully homomorphic encryption over the integers. In *Advances in Cryptology– EUROCRYPT '10* (pp. 24–43). doi:10.1007/978-3-642-38348-9_20